\documentclass[12pt]{article}
\usepackage{epsfig}
\usepackage{graphicx}

\usepackage{epsfig}
\usepackage{graphicx}
\usepackage{bbm}

\newcommand{\nc}{\newcommand}
\nc{\beq}{\begin{equation}}  \nc{\eeq}{\end{equation}}
\nc{\bea}{\begin{eqnarray}}  \nc{\eea}{\end{eqnarray}}
\nc{\baa}{\begin{array}}     \nc{\eaa}{\end{array}}
\nc{\bit}{\begin{itemize}}   \nc{\eit}{\end{itemize}}
\nc{\ben}{\begin{enumerate}} \nc{\een}{\end{enumerate}}
\nc{\bce}{\begin{center}}    \nc{\ece}{\end{center}}
\nc{\bpm}{\begin{pmatrix}}   \nc{\epm}{\end{pmatrix}}
\nc{\bvt}{\begin{verbatim}}  \nc{\evt}{\end{verbatim}}

\def\inv#1{\frac1{#1}}
\def\half{\frac12}
\def\vev{vacuum expectation value}

\def\su#1{{SU(#1)}}
\def\ui{U(1)}
\def\vevof#1{\left\langle #1 \right\rangle}

\def\rcal{{\cal R}}

\def\Chi{X}
\def\gl{g_L}
\def\gr{g_R}
\def\thw{\theta_{W}}
\def\tw{t_{W}}
\def\cw{c_{W}}
\def\mz{m_{Z}}
\def\mzp{m_{Z'}}
\def\dz{\Delta_{Z}}
\def\mwl{m_{W_L}}
\def\mwr{m_{W_R}}
\def\dw{\Delta_{W}}
\def\hc{\hbox{H.c.}}
\def\tr{\hbox{Tr}}
\def\pt{\tilde\Phi}
\def\ptd{\pt^\dagger}
\def\pd{\Phi^\dagger}
\def\cd{\Chi^\dagger}
\def\pr{\Phi_R}
\def\prd{\pr^\dagger}
\def\prt{\tilde\pr}
\def\prtd{\prt^\dagger}
\def\qwe{f'}
\def\asd{u}
\def\zxc{z}
\def\ca{c_\alpha}
\def\sa{s_\alpha}
\def\ta{\tan\alpha\;}
\def\sta{s_{2\alpha}}

\catcode`\@=11
\def\lsim{\mathrel{\mathpalette\@versim<}}
\def\gsim{\mathrel{\mathpalette\@versim>}}
\def\@versim#1#2{\vcenter{\offinterlineskip
\ialign{$\m@th#1\hfil##\hfil$\crcr#2\crcr\sim\crcr } }}
\catcode`\@=12

\catcode`\@=11
\def\lsim{\mathrel{\mathpalette\@versim<}}
\def\gsim{\mathrel{\mathpalette\@versim>}}
\def\@versim#1#2{\vcenter{\offinterlineskip
\ialign{$\m@th#1\hfil##\hfil$\crcr#2\crcr\sim\crcr } }}
\catcode`\@=12

\catcode`@=12
\begin{document}
\thispagestyle{empty}
\begin{flushright}
UCRHEP-T473\\
DCP-09-03 \\
October 2009\
\end{flushright}
\begin{center}
{\LARGE \bf Asymmetric Higgs Sector and\\ Neutrino Mass in an $SU(2)_R$ Model\\}
\vspace{0.5in}
{\bf Alfredo Aranda$^{a,b}$, J. Lorenzo Diaz-Cruz$^{b,c}$, Ernest Ma$^d$,\\
Roberto Noriega$^{b,e}$, and Jose Wudka$^d$\\}
\vspace{0.2in}
{\sl $^a$ Facultad de Ciencias - CUICBAS, Universidad de Colima, \\
Colima, Col. M\'exico\\}
\vspace{0.1in}
{\sl $^b$ Dual CP Institute of High Energy Physics, M\'exico\\}
\vspace{0.1in}
{\sl $^c$ C.A. de Particulas, Campos y Relatividad, \\
FCFM-BUAP, Puebla, Pue., Mexico\\}
\vspace{0.1in}
{\sl $^d$ Department of Physics and Astronomy, University of California,\\
Riverside, California 92521, USA\\}
\vspace{0.1in}
{\sl $^e$ CIMA Universidad Aut\'onoma del Estado de Hidalgo, Pachuca,
Hgo. M\'exico\\}
\end{center}
\begin{abstract}\
The asymmetric Higgs sector of one $SU(2)_L \times SU(2)_R$ bidoublet
$(\phi_1^0,\phi_1^-;\phi_2^+,\phi_2^0)$ and one $SU(2)_R$ doublet [but no
$SU(2)_L$ doublet] is considered in a nonsupersymmetric left-right extension
of the standard model (SM) of particle interactions.  The inverse seesaw
mechanism for neutrino mass is naturally implemented with the addition of
fermion singlets, allowing thereby the possibility of breaking $SU(2)_R$
at the TeV scale.  Flavor-changing neutral Higgs couplings to quarks are
studied in two scenarios, where the $SU(2)_R$ charged-current mixing
matrix is given either by $V_R = V_{CKM}$ (scenario I)
or $V_R = 1$ (scenario II).  We consider the bounds on these scalar
particle masses from $K-\bar{K}$ and $B-\bar{B}$ mixing, as well as
$b \to s \gamma$.  We find that, whereas in scenario I, they are of order
10 TeV, as in other left-right models, they may be well below 1 TeV in
scenario II, thus allowing them to be within reach of detection at the
forthcoming Large Hadron Collider (LHC).
\end{abstract}

\newpage
\baselineskip 24pt

\section{Introduction}

In the nonsupersymmetric $SU(3)_C \times SU(2)_L \times SU(2)_R \times
U(1)_{B-L}$ extension of the standard $SU(3)_C \times SU(2)_L \times U(1)_Y$
model (SM) of particle interactions, the Higgs sector must be enlarged from
the one $SU(2)_L$ scalar doublet of the SM.  There are several ways to do
this, as discussed comprehensively in Ref.~\cite{m04}.  In the canonical
approach, a Higgs triplet is used to break $SU(2)_R$ at a large scale, and
$\nu_R$ gets a large Majorana mass.  A Higgs bidoublet is then added to break
$SU(2)_L$, and all fermions obtain Dirac masses, with $\nu_L$ getting a small
seesaw mass. In this scenario, the $SU(2)_R$ breaking scale is presumably
beyond the reach of present accelerators, such as the Large Hadron Collider
(LHC). Even if we try to lower this scale, the canonical model has severe
difficulties with flavor-changing neutral currents, in contradiction
with what is experimentally observed.

The purpose of this paper is to elaborate on a simple alternative~\cite{m04},
where the $SU(2)_R$ breaking scale may be lowered to 1 TeV, using the
inverse seesaw mechanism for neutrino mass~\cite{ww83,mv86,m87,m09-1}.
We choose a Higgs sector which contains only one $SU(2)_L \times SU(2)_R$
bidoublet and one  $SU(2)_R$ doublet [but no $SU(2)_L$ doublet]. Of course,
flavor-changing neutral Higgs couplings are still unavoidable.  However,
as we show in this paper, a scenario exists where they are sufficiently
suppressed.  Since the $SU(2)_R$ charged-current mixing matrix is unknown,
we consider two scenarios, where it is given either by $V_R = V_{CKM}$
(scenario I) or $V_R = 1$ (scenario II).  We consider the bounds on the
corresponding scalar particle masses from $K-\bar{K}$ and $B-\bar{B}$ mixing,
as well as $b \to s \gamma$.  We find that, whereas in scenario I, they are
of order 10 TeV, as in other left-right models, they may be well below 1 TeV
in scenario II, thus allowing them to be within reach of detection at the
forthcoming Large Hadron Collider (LHC).

\section{Asymmetric Left-Right Model}

\subsection{Particle content and neutrino mass}

The fermion content of the minimal $SU(3)_C \times SU(2)_L \times SU(2)_R
\times U(1)_{B-L}$ gauge model is well-known, i.e.
\begin{eqnarray}
&& \psi_L = \pmatrix{\nu_e \cr e}_L \sim (1,2,1,-1/2), ~~~
\psi_R = \pmatrix{\nu_e \cr e}_R \sim (1,1,2,-1/2), \\
&& Q_L = \pmatrix{u \cr d}_L \sim (3,2,1,1/6), ~~~
Q_R = \pmatrix{u \cr d}_R \sim (3,1,2,1/6),
\end{eqnarray}
where the $U(1)$ charge is normalized to $(B-L)/2$ so that the electric
charge is given by $Q = I_{3L} + I_{3R} + (B-L)/2$.  Here a neutral fermion
singlet
\begin{equation}
S_L \sim (1,1,1,0)
\end{equation}
is also added per family, which will have important implications for the
neutrino masses, as shown below.

To obtain masses for the quarks and leptons, a Higgs bidoublet
\begin{equation}
\Phi = \pmatrix{\phi_1^0 & \phi_2^+ \cr \phi_1^- & \phi_2^0} \sim (1,2,2,0)
\end{equation}
is needed.  In a nonsupersymmetric model, which is being considered
here, the dual of $\Phi$, i.e.
\begin{equation}
\tilde{\Phi} = \sigma_2 \Phi^* \sigma_2 = \pmatrix{\bar{\phi}_2^0 & -\phi_1^+
\cr -\phi_2^- & \bar{\phi}_1^0} \sim (1,2,2,0)
\end{equation}
must also be used. To break $SU(2)_R \times U(1)_{B-L}$ to $U(1)_Y$, an
$SU(2)_R$ Higgs doublet
\begin{equation}
\Phi_R = \pmatrix{\phi_R^+ \cr \phi_R^0}
\end{equation}
is added, which also links $\bar{\nu}_R$ with $S_L$ to form a Dirac mass $m_R$.
Since $S_L$ is a gauge singlet, it is also allowed to have a Majorana mass
$m_S$; hence the $3 \times 3$ neutrino mass matrix spanning
$(\bar{\nu}_L, \nu_R, \bar{S}_L)$ is of the form
\begin{equation}
{\cal M}_{\nu,S} = \pmatrix{0 & m_D & 0 \cr m_D & 0 & m_R \cr 0 & m_R & m_S},
\end{equation}
where $m_D$ is the usual Dirac mass linking $\bar{\nu}_L$ to $\nu_R$ through
$\langle \phi_1^0 \rangle$ and $\langle \bar{\phi}_2^0 \rangle$.  A quick
look at the above shows clearly that if $m_S=0$, then lepton number is
conserved with a linear combination of $\nu_L$ and $S_L$ forming a Dirac
fermion with $\nu_R$, and the orthogonal combination is exactly massless.
This means that it is natural for $m_S$ to be small, thereby triggering the
inverse seesaw mechanism, resulting in
\begin{equation}
m_\nu \simeq {m_D^2 m_S \over m_R^2}.
\end{equation}
Note that there is no entry in Eq.~(7) linking $\nu_L$ and $S_L$ because the
$SU(2)_L$ Higgs doublet is absent.  This is important for the validity of
Eq.~(8).  Instead of the canonical seesaw formula $m_\nu \simeq -m_D^2/m_R$,
which is small if $m_R$ is large, Eq.~(8) lets $m_\nu$ be small
if $m_S$ is small, even if $m_R$ is not too large.  Thus the inverse
seesaw mechanism is suitable for bringing down the scale of $SU(2)_R$
breaking to 1 TeV, with verifiable phenomenology at the LHC.  Note
also that the mixing of $\nu_L$ with $S_L$ is of order
$m_D/m_R$ which may now be nonnegligible and results in deviations from
unitarity \cite{m09-2} of the neutrino mixing matrix.

\subsection{Higgs sector}

The most general Higgs potential consisting of $\Phi_R$, $\Phi$, and
$\tilde \Phi$ is given by
\begin{eqnarray}
V &=& m_R^2 \Phi_R^\dagger \Phi_R + m^2 Tr(\Phi^\dagger \Phi) + {1 \over 2}
\mu^2 Tr(\Phi^\dagger \tilde \Phi + \tilde{\Phi}^\dagger \Phi) \nonumber \\
&+& {1 \over 2} \lambda_R (\Phi_R^\dagger \Phi_R)^2 + {1 \over 2} \lambda_1
[Tr(\Phi^\dagger \Phi)]^2 + {1 \over 2} \lambda_2 Tr(\Phi^\dagger \Phi
\Phi^\dagger \Phi) \nonumber \\ &+& {1 \over 8} \lambda_3 \{ [Tr(\Phi^\dagger
\tilde \Phi)]^2 + [Tr(\tilde{\Phi}^\dagger \Phi)]^2 \} + {1 \over 2} \lambda_4
[Tr(\Phi^\dagger \Phi)][Tr(\Phi^\dagger \tilde \Phi + \tilde{\Phi}^\dagger \Phi)]
\nonumber \\ &+& f_1 \Phi_R^\dagger (\tilde{\Phi}^\dagger \tilde{\Phi}) \Phi_R +
f_2 \Phi_R^\dagger (\Phi^\dagger \Phi) \Phi_R + f_3 \Phi_R^\dagger (\Phi^\dagger
\tilde{\Phi} + \tilde{\Phi}^\dagger \Phi) \Phi_R,
\end{eqnarray}
where all parameters have been chosen real for simplicity. Let $\langle
\phi_R^0 \rangle = v_R$ and $\langle \phi_{1,2}^0 \rangle = v_{1,2}$, then
the minimum of $V$ is given by
\begin{eqnarray}
V_0 &=& m_R^2 v_R^2 + m^2 (v_1^2 + v_2^2) + 2 \mu^2 v_1 v_2 + {1 \over 2}
\lambda_R v_R^4 + {1 \over 2} \lambda_1 (v_1^2+v_2^2)^2 + {1 \over 2} \lambda_2
(v_1^4 + v_2^4) \nonumber \\ &+& \lambda_3 v_1^2 v_2^2 + 2 \lambda_4 (v_1^2 +
v_2^2) v_1 v_2 + f_1 v_1^2 v_R^2 + f_2 v_2^2 v_R^2 + 2 f_3 v_1 v_2 v_R^2,
\end{eqnarray}
where $v_R$ and $v_{1,2}$ satisfy
\begin{equation}
v_R (m_R^2 + \lambda_R v_R^2 + f_1 v_1^2 + f_2 v_2^2 + 2 f_3 v_1 v_2) = 0,
\end{equation}
\begin{equation}
v_1[m^2 + f_1 v_R^2 + (\lambda_1 + \lambda_2) v_1^2 + (\lambda_1 +
\lambda_3) v_2^2 + 3 \lambda_4 v_1 v_2] + v_2 (\mu^2 + f_3 v_R^2 + \lambda_4
v_2^2) = 0,
\end{equation}
\begin{equation}
v_2[m^2 + f_2 v_R^2 + (\lambda_1 + \lambda_2) v_2^2 + (\lambda_1 +
\lambda_3) v_1^2 + 3 \lambda_4 v_1 v_2] + v_1 (\mu^2 + f_3 v_R^2 + \lambda_4
v_1^2) = 0.
\end{equation}
A solution exists where $v_2 \ll v_1$, i.e.
\begin{equation}
v_2 \simeq {-(\mu^2 + f_3 v_R^2 + \lambda_4 v_1^2) v_1 \over m^2 + f_2 v_R^2 +
(\lambda_1 + \lambda_3) v_1^2},
\end{equation}
with
\begin{equation}
v_1^2 = {m_R^2 f_1 - m^2 \lambda_R \over \lambda_R ( \lambda_1+\lambda_2)
- f_1^2}, ~~~ v_R^2 = {-m_R^2 - f_1 v_1^2 \over \lambda_R}.
\end{equation}
Fine tuning is of course unavoidable.  In the limit $v_2=0$, the physical
Higgs bosons are $\phi_2^\pm$ and $Im\phi_2^0$ with masses squared
\begin{equation}
m^2(\phi_2^\pm) = (f_2-f_1) v_R^2, ~~~ m^2(Im\phi_2^0) =
(f_2-f_1) v_R^2 - (\lambda_2 + \lambda_3) v_1^2,
\end{equation}
and three linear combinations of $Re\phi_1^0$, $Re\phi_R^0$,
$Re\phi_2^0$, with mass-squared matrix
\begin{equation}
\label{M_scalar}
{\cal M}^2 = \pmatrix{2 (\lambda_1 + \lambda_2) v_1^2 & 2 f_1 v_1 v_R &
2 \lambda_4 v_1^2 \cr 2 f_1 v_1 v_R & 2 \lambda_R v_R^2 & 2 f_3 v_1 v_R \cr
2 \lambda_4 v_1^2 & 2 f_3 v_1 v_R & (f_2 - f_1) v_R^2 - (\lambda_2 - \lambda_3)
v_1^2}.
\end{equation}

\subsection{Gauge bosons}

The structure of the scalar sector leads in general to both $W_L - W_R$ and
$Z - Z'$ mixing.  The former vanishes in the limit $ v_2 \to 0 $  and so will
be suppressed for the above choice of vacuum  expecation values. In contrast,
the $Z-Z'$ mixing term is proportional to $v_1^2 $, which is unacceptably
large. To cancel this contribution, a simple possibility is to add a Higgs
bidoublet $X \sim (1,2,2,-1)$ with vacuum expectation value $v_3$.  In that
case, the choice $v_3^2/v_1^2 = 1 - 2 \sin^2 \theta_W$ (for $g_L=g_R$) will
lead to zero mixing at tree level; details are given in the Appendix.  Note
that $X$ will not affect the $\rho$ parameter (at tree level) in precision
electroweak measurements, nor will it contribute to quark or lepton masses.
In particular, it does not link $\nu_L$ with $S_L$ in Eq.~(7), otherwise the
inverse seesaw mechanism would be invalidated.  The present experimental
limits on $W_R$ and $Z'$ are respectively 715 and 860 GeV.

\section{Flavor-Changing Processes from Neutral Higgs Couplings}

\subsection{General structure}

Since both $\Phi$ and $\tilde{\Phi}$ couple to the quarks and leptons,
flavor-changing interactions through the exchange of neutral Higgs
scalars are unavoidable. The question is whether they can be
suppressed \cite{lrmohapatretal}.
Consider the Yukawa terms
\begin{equation}
(h^u_{ij} \phi_1^0 + h^d_{ij} \bar{\phi}_2^0) \bar{u}_{iL} u_{jR} +
(h^u_{ij} \phi_2^0 + h^d_{ij} \bar{\phi}_1^0) \bar{d}_{iL} d_{jR}.
\end{equation}
In the limit $v_2=0$, both $up$ and $down$ quark masses come from only $v_1$.
Hence
\begin{equation}
h^u_{ij} v_1 = U_L \pmatrix{m_u & 0 & 0 \cr 0 & m_c & 0 \cr 0 & 0 & m_t}
U_R^\dagger, ~~~ h^d_{ij} v_1 = D_L \pmatrix{m_d & 0 & 0 \cr 0 & m_s & 0
\cr 0 & 0 & m_b} D_R^\dagger,
\end{equation}
where $U_{L,R}$ and $D_{L,R}$ are unitary matrices, with
\begin{equation}
U_L^\dagger D_L = V_{CKM}, ~~~ U_R^\dagger D_R = V_R,
\end{equation}
being the quark mixing matrix for the known left-handed charged currents
and that for their unknown right-handed counterparts.  This means that in the
basis of quark mass eigenstates, the structure of flavor-changing neutral
currents through scalar exchange is determined, i.e.
\begin{equation}
{Re \phi_1^0 \over v_1} \pmatrix{m_u & 0 & 0 \cr 0 & m_c & 0 \cr 0 & 0 & m_t}
+ {\bar{\phi}_2^0 \over v_1} ~V_{CKM} \pmatrix{m_d & 0 & 0 \cr 0 & m_s & 0 \cr
0 & 0 & m_b} V_R^\dagger
\end{equation}
for the $up$ quarks, and
\begin{equation}
{Re \phi_1^0 \over v_1} \pmatrix{m_d & 0 & 0 \cr 0 & m_s & 0 \cr 0 & 0 & m_b}
+ {{\phi}_2^0 \over v_1} ~V_{CKM}^\dagger \pmatrix{m_u & 0 & 0 \cr 0 & m_c & 0
\cr 0 & 0 & m_t} V_R
\end{equation}
for the $down$ quarks.  Hence $Re\phi_1^0$ behaves as the SM Higgs
boson, and at tree level, all flavor-changing effects come from
$\phi_2^0$, whereas in one loop, there are also contributions from
$(\phi_2^+,\phi_2^0)$.  Note that for $v_1^2 \ll v_R^2$, this electroweak
doublet has the common mass of $\sqrt{f_2-f_1} v_R$.


In the lepton sector, the analog of $V_{CKM}$ is unknown because the
neutrino mass matrix depends on $m_D$, $m_R$, and $m_S$.  In fact, we
could choose $m_D$ to be diagonal in the $(e,\mu,\tau)$ basis and still
have the freedom to obtain the observed neutrino mixing matrix from
$m_R$ and $m_S$.  In that case, $\phi_2^0$ would have no flavor-changing
leptonic interactions.

\subsection{$K-\bar{K}$ and $B-\bar{B}$ mixing}

We now apply Eq.~(22) to $K-\bar{K}$ and $B-\bar{B}$ mixing.  In the
two scenarios I and II considered for the $V_R$ matrix mentioned in
the Introduction, the $\phi_2^0$ couplings are of the form:
\begin{eqnarray}
&{\rm (I)}& V_R = V_{CKM}~:~~~ {\phi_2^0 \over v_1} \bar{d}_{iL} d_{jR}
\sum_k m_{u_k} V^*_{u_k d_i} V_{u_k, d_j} + H.c.\\
&{\rm (II)}& V_R = 1~:~~~ {\phi_2^0 \over v_1} \bar{d}_{iL} d_{jR}
~m_{u_j} V^*_{u_j d_i} + H.c.
\end{eqnarray}

We use the formulae presented in Ref.~\cite{ADandAP03}.  The mass difference
of a neutral meson and its antiparticle is written in terms of its SM
and other contributions:
\begin{equation}
\Delta M_X = (\Delta M)_{X,SM} + (\Delta M)_{X,New}
\end{equation}
where $\Delta M_X = \, \Delta M_K,\,\Delta M_{B_d}, \, \Delta M_{B_s}$, and
$(\Delta M)_{X,SM}$ denotes the SM (one-loop) contribution, and
$(\Delta M)_{X,New}$ is everything else.  In our case, the latter comes from
the flavor-changing $\phi^0_2$ couplings. The resulting expression
for the mass difference is then given by
\begin{equation}
\label{DeltaM}
(\Delta M)_{X,New}= \frac{G^2_F \, M^2_W}{6 \pi^2}\, S_X
          \left[ \bar{P}^{LR}_2 {C}^{LR}_2 + \bar{P}^{SLL}_1 ( {C}^{SLL}_1
+ {C}^{SRR}_1) \right]
\end{equation}
where the constant $S_X$ includes strong-interaction effects, and the
coefficients $P$ include next-to-leading QCD corrections, while the
functions $C$ denote the Wilson coefficients of the OPE expansion for
the relevant hadronic matrix elements.

Let us consider first case (I) of our model, i.e. $V_R = V_{CKM}$.  Here the
Wilson coefficients  $C^{SLL}_1, C^{SRR}_1$ are equal:
\begin{equation}
\label{wilsonA}
 C^{SLL}_1 = C^{SRR}_1 = \frac{16 \pi^2}{G^2_F M^2_W}
                                   \left( \frac{r^{LL}_X }{ v_1} \right)^2
                                   \left[ \frac{1}{m_{Re\phi^0_2}^2}
-\frac{1}{m_{Im\phi^0_2}^2} \right],
\end{equation}
and suppressed because the mass difference between $Re\phi_2^0$ and
$Im\phi_2^0$ is small compared to their sum, whereas $C_2^{LR}$ is of the form:
\begin{equation}
\label{wilsonB}
 C^{LR}_2 = \frac{16 \pi^2}{G^2_F M^2_W}
                                   \left( \frac{r^{LR}_X }{ v_1} \right)^2
                                   \left[ \frac{1}{m_{Re\phi^0_2}^2}
+ \frac{1}{m_{Im\phi^0_2}^2} \right],
\end{equation}
which has no such suppression.  In case (I), the various $r$'s in each system
are also the same: $r^{LR}_X=r^{LL}_X = r^{RR}_X= r_X$,  where
\begin{eqnarray}
r_K &=& m_u V_{ud} V_{us} + m_c V_{cd} V_{cs} + m_t V_{td} V_{ts}, \\
r_{B_d} &=& m_u V_{ud} V_{ub} + m_c V_{cd} V_{cb} + m_t V_{td} V_{tb}, \\
r_{B_s} &=& m_u V_{us} V_{ub} + m_c V_{cs} V_{cb} + m_t V_{ts} V_{tb}.
\end{eqnarray}
We have also assumed for simplicity that all the $V_{CKM}$ entries
are real.

Obviously there are large contributions coming from those terms proportional
to $m_t$ or $m_c$.  However, there is also a natural suppression for
the $C^{LL}$ and $C^{RR}$ Wilson coefficients, because their contributions
are proportional to the effective $\langle \phi_2^0 \phi_2^0 \rangle$
propagator, i.e. $m^{-2}(Re\phi_2^0) - m^{-2}(Im\phi_2^0)$.  Whereas
$Im\phi^0_2$ is a mass eigenstate, $Re\phi^0_2$ is not, but if $f_3$ and
$\lambda_4$ are small in Eq.~(\ref{M_scalar}), then it is approximately so,
and their combined contribution for $v_1^2 \ll v_R^2$ is naturally suppressed,
i.e.
\begin{equation}
\label{propagator}
\frac{1}{(f_2 - f_1) v_R^2 - (\lambda_2 - \lambda_3)
v_1^2 } - \frac{1}{(f_2 - f_1) v_R^2 - (\lambda_2 + \lambda_3)
v_1^2 } \simeq \frac{-2 \lambda_3 v^2_1}{(f_2 - f_1)^2 v^4_R}.
\end{equation}
This suppression persists even if $f_3$ and $\lambda_4$ are not neglected.
We simply replace $\lambda_3$ by
\begin{equation}
\lambda_3 + {2 f_3 \lambda_4 f_1 - f_3^2(\lambda_1+\lambda_2) -
\lambda_4^2 \lambda_R \over \lambda_R(\lambda_1+\lambda_2) - f_1^2}.
\end{equation}
This feature of our model would allow $v_R$ to be at the TeV scale, without
running into conflict with present data on $K-\bar{K}$ and $B-\bar{B}$ mixing
as far as $C^{LL}$ and $C^{RR}$ are concerned. Unfortunately, this suppression
does not work for $C^{LR}$, which is proportional to $m^{-2}(Re\phi_2^0) +
m^{-2}(Im\phi_2^0)$.  However, as we show below in case (II), the $C^{LR}$
coefficients are further suppressed by light quark masses in the $r$'s,
which allows $\phi_2^0$ to be lighter than 1 TeV.

In case (II), i.e. $V_R=1$, the $r$ values are related by
$(r^{LR}_X)^2=r^{LL}_X r^{RR}_X$, with
\begin{eqnarray}
&& r^{LL}_K = m_c V_{cd},     ~~~  r^{RR}_K = m_u V_{us}, \\
&& r^{LL}_{B_d} = m_t V_{td}, ~~~  r^{RR}_{B_d} = m_u V_{ub}, \\
&& r^{LL}_{B_s} = m_t V_{ts}, ~~~  r^{RR}_{B_s} = m_c V_{cb}.
\end{eqnarray}
From the above, it is clear that whereas $C^{LR}$ is not suppressed by
Eq.~(\ref{propagator}), it is much smaller than what it is in case (I),
because of the smallness of $r^{LR}$.

As mentioned in Ref.~\cite{ADandAP03}, there are large theoretical
uncertainties associated with these expressions.  To make an estimate, we
simply require the absolute value of the contribution of new physics to be
less than the corresponding experimental value.  In what follows we shall
obtain bounds for the combination of parameters:
$1/\Delta^2 = m^{-2}(Re\phi_2^0) - m^{-2}(Im\phi_2^0)$ and
$1/ \Sigma^2= m^{-2}(Re\phi_2^0) + m^{-2}(Im\phi_2^0)$.
Let us define: $\Sigma^2= m^{2}_2/2$ and
$1/\Delta^2 = \delta^2/ m^{4}_2$, where $m_2$ is the approximate mass of
$Re(\phi_2^0)$ or $Im(\phi_2^0)$, and $\delta^2$ is a meassure of the
splitting between their squared masses.

Using  Eqs.~(\ref{wilsonA}) and (\ref{wilsonB}), we obtain the following
general expression:
\begin{equation}
\left( \frac{r^{LR}_X }{ v_1} \right)^2  \frac{2 P^{LR}_2}{m^2_2} +
\left[ \left( \frac{r^{LL}_X }{ v_1} \right)^2 + \left( \frac{r^{RR}_X }
{ v_1} \right)^2
\right]
 \frac{P^{SLL}_1 \delta^2 }{m^4_2}  =
\frac{ 3 }{ 8 S_X } \Delta M_{X}^{Exp}.
\end{equation}
For the $K-\bar{K}$ system, $S_K= m_k  \, F^2_K \, \eta_2 \,\hat{B}_K$,
with $F_K=160$ MeV, $m_K=498$ MeV, $\eta_2=0.57$ and $\hat{B}_K=0.85$.
At the scale $\mu=2$ GeV, $\bar{P}^{LR}_2=30.6$, $\bar{P}^{SLL}_1=-9.3$,
$\Delta M_K^{Exp}= 3.48 \times 10^{-12}$ MeV.  Notice that in case (I),
both $P^{LR}_2$ and $C^{LR}_2$ dominate over the $LL$ and $RR$ contributions.
Threfore the resulting bound is not sensitive to the parameter $\delta$,
and the bound on $m_2$ is given by
\begin{equation}
m_2 \geq 25 \, \, {\rm TeV}
\end{equation}
For the $B-\bar{B}$ systems, we take the corresponding parameters
from the Particle Data Group~\cite{PDG-09}, so that for $(B_d, B_s)$:
\begin{equation}
m_2 \geq 12 (11) \, \, {\rm TeV}
\end{equation}
These results are in agreement with \cite{lrmohapatretal}.

In case (II), if we take $\delta=0$ (i.e. only the $LR$
contribution), we obtain a much smaller bound for the $K$ system,
i.e. $m_2 \geq 1.1$ TeV. However, for the $B_d$, and $B_s$
systems, the same procedure yields the bounds $m_2 \geq 60 (900)$
GeV, respectively.  Thus for the $B_d$ system, it seems more
appropriate to consider $\delta \neq 0$, in which case the bound
becomes: ${m^2_2}/{\delta} \geq 3.7$ TeV.

\subsection{$b \to s \gamma$}

\begin{figure}[htb]
\centering
\includegraphics[width=13cm]{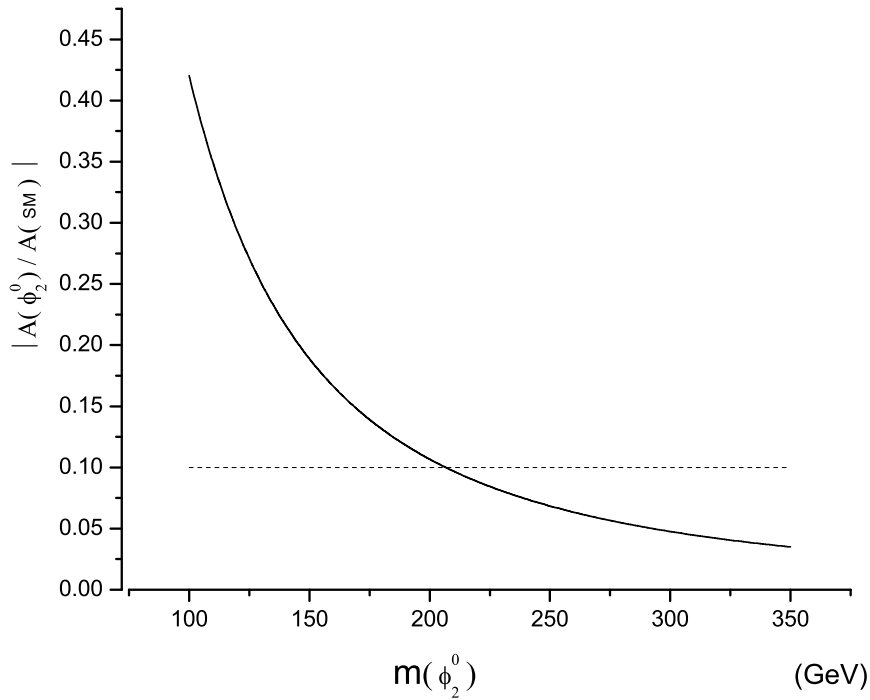}
\caption{ Plot of $|A_{\phi_2^0}/A_{SM}|$ vs $m_{\phi_2^0}$.}
\label{fig}
\end{figure}

To evaluate the contribution of $(\phi_2^+,\phi_2^0)$ to $b \to s \gamma$,
we consider the relevant terms in Eq.~(22).  For case (I), i.e.
$V_R = V_{CKM}$, the important ones are
\begin{equation}
{m_t \over v_1} |V_{tb}|^2 (\phi_2^0 \bar{b}_L + \phi_2^+ \bar{t}_L) b_R +
{m_t \over v_1} V_{ts}^*V_{tb} (\phi_2^0 \bar{s}_L + \phi_2^+ \bar{c}_L) b_R +
{m_t \over v_1} V_{tb}^*V_{ts} (\phi_2^0 \bar{b}_L + \phi_2^+ \bar{t}_L) s_R
+H.c.
\end{equation}
For case (II), i.e. $V_R = 1$, they are
\begin{equation}
{m_t \over v_1} V_{tb}^* (\phi_2^0 \bar{b}_L + \phi_2^+ \bar{t}_L) b_R +
{m_t \over v_1} V_{ts}^* (\phi_2^0 \bar{s}_L + \phi_2^+ \bar{c}_L) b_R + H.c.
\end{equation}
The SM contribution (from $W$ exchange) is of the form $\bar{s}_L
\sigma_{\mu \nu} b_R$ which is classified~\cite{Bertolini:1990if} as $O_7$.
Using the above interactions, there is only one such contribution coming
from $\phi_2^0$ exchange, i.e. $\bar{s}_L b_R$ and $\bar{b}_R b_L$, which
is proportional to $V_{ts}^* m_t^2/v_1^2$ in both cases (I) and (II),
assuming that $V_{tb}=1$, which is of course a very good
approximation.  In contrast to the usual two-Higgs-doublet model, the
$\phi_2^+$ contribution is suppressed here because it is proportional to
$m_b$.  As for $O'_7$, i.e. $\bar{s}_R \sigma_{\mu \nu} b_L$,
both $\phi_2^0$ and $\phi_2^+$ have contributions proportional to
$V_{ts}^* m_t^2/v_1^2$, but since the $b \to s \gamma$ rate is
proportional to
\begin{equation}
|C_7|^2 + |C'_7|^2 = |A_{SM} + A_{\phi_2^0}|^2 + |A'_{\phi_2^0} + A'_{\phi_2^+}|^2,
\end{equation}
the latter can be safely ignored.  Using Ref.~\cite{Bertolini:1990if}, we find
\begin{eqnarray}
A_{SM} &\sim & \frac{3 m^2_t}{m^2_{W}} [ \frac{2}{3} F_1(x_t) +F_2(x_t) ]   \\
A_{\phi^0_2} &\sim& \frac{m^2_t}{m^2_{\phi^0_2}} [ -\frac{1}{3} F_1(x_b)]
\end{eqnarray}
where $x_t=m^2_t/m^2_W$, $x_b=m^2_b/m^2_{\phi^0_2}$,
and the functions $F_{1,2}$ are given by
\begin{eqnarray}
F_1(x) = {1 \over 12(x-1)^4} (x^3 - 6x^2 + 3x + 2 + 6 x \ln x), \\
F_2(x) = {1 \over 12(x-1)^4} (2x^3 + 3 x^2 - 6x + 1 - 6x^2 \ln x).
\end{eqnarray}
We now require the amplitude ratio $|A_{\phi_2^0}/A_{SM}|$ to be less than
10\%, so that it is well within the experimental accuracy.  This translates
to an estimated lower bound for $m_{\phi_2^0}$ of about 200 GeV, as shown in
Fig.~\ref{fig}.

\section{Conclusion}

We have studied in this paper a simple nonsupersymmetric left-right
extension of the standard model.  The asymmetric Higgs sector of this
model consists of one $SU(2)_L \times SU(2)_R$ bidoublet and one $SU(2)_R$
doublet [but no $SU(2)_L$ doublet].  With the addition of neutral fermion
singlets, the inverse seesaw mechanism for neutrino mass is naturally
implemented, suggesting that the $SU(2)_R$ breaking scale may be lowered
to 1 TeV. We then analyzed the unavoidable problem of flavor-changing
couplings of the neutral Higgs bosons of this model and showed that in
the limit of $v_2 = \langle \phi_2^0 \rangle = 0$, these effects are
naturally suppressed in case (II) ($V_R = 1$) [but not in case (I)
($V_R = V_{CKM}$), which has the same constraint as other left-right
models that the $SU(2)_R$ breaking scale is above 10 TeV.]   From
$K-\bar{K}$ and $B-\bar{B}$ mixing, we find $v_R = \langle \phi_R^0 \rangle$
to be consistent with less than about 1 TeV in case (II).   From $b \to s
\gamma$, we find $m_{\phi_2^0}$ to be above 200 GeV.  The new particles of
this model, i.e. $W_R^\pm$, $Z'$, the heavy pseudo-Dirac neutral fermion of
mass $m_R$ from the pairing $S_L$ with $\nu_R$, and the heavy Higgs particles
$Re\phi_R^0$ and $(\phi_2^+,\phi_2^0)$, are all consistent with having
masses below 1 TeV in case (II) and are potentially observable at the LHC.

\section*{Acknowledgments}

This work was supported in part by the U.~S.~Department of Energy under
Grant No.~DE-FG03-94ER40837, by UC-MEXUS, and by CONACYT. We also
thank M. Holthauseen for comments and pointing out a discrepancy in
the first version of this paper.

\section*{Appendix}

With only the bidoublet $\Phi$, our model exhibits an unavoidable $Z - Z' $
mixing term proportional to $v_1^2$, implying thus a very large value of $v_R$.
This can be remedied by enlarging the scalar sector through the addition
of another bidoublet $ \Chi $ of  $(B-L)/2 = -1$,
\beq
\Chi = \bpm{\chi_1^- & \chi_2^0 \cr \chi_1^{--} & \chi_2^-}\epm
\sim (1,2,2,-1) \,,
\eeq
and its corresponding dual $ \tilde\Chi = \sigma_2 \Chi^* \sigma_2 $.
We list in this appendix the modifications resulting from its addition.

\paragraph{Vector-boson masses:}
Let the neutral components of $\Phi$, $\Phi_R$, and $\Chi$ acquire
vacuum expectation values
\beq
\vevof{\phi_{1,2}^0} = v_{1,2} \,, \quad
\vevof{\phi_R^0} = v_R \,, \quad
\vevof{\chi_2^0} = v_3 \,,
\eeq
and denote the neutral gauge bosons associated with $\su2_{L,R}$ by
$W^0_{L,R}$ and the $\ui$ gauge boson by $B$, with $ g_{L,R}$
and $g'$ the gauge couplings for $\su2_{L,R} $ and $ \ui$
respectively.  The resulting mass-squared matrix in the $(W_R^0, W_L^0, B)$
basis is then given by
\beq
{\cal M}^2 = 2 \bpm{
\gr^2 ( v_1^2 + v_2^2 + v_3^2 + v_R^2) &
 - \gl \gr  ( v_1^2 + v_2^2 - v_3^2 ) &
- g' \gr (v_R^2 + 2 v_3^2) \cr
 - \gl \gr  ( v_1^2 + v_2^2 - v_3^2 ) &
\gl^2 ( v_1^2 + v_2^2 + v_3^2 ) &
- 2 g' \gl v_3^2 \cr
- g' \gr (v_R^2 + 2 v_3^2) &
- 2 g' \gl v_3^2 &
- g' \gr (v_R^2 + 4 v_3^2) } \epm \,.
\eeq
The photon $A$, the neutral gauge boson $Z$ of the SM, and the new $Z'$
are then linear combinations, determined according to
\beq
\bpm{ W_R^0 \cr W_L^0 \cr B }\epm = \rcal \bpm{ A \cr Z \cr Z' }\epm;
\quad
\rcal =e \bpm{
1/\gr & \tw/\gr & -1/(g' \cw) \cr
1/\gl & -1/(\tw \gl ) & 0 \cr
1/g' & \tw/g' & 1/(\gr \cw) \,
} \epm ,
\eeq
where $ \cw = \cos\thw,~ \tw = \tan\thw $ and
the weak-mixing angle $\thw$ and the proton charge $e$
are defined by
\beq
\tan\thw = \frac{ g' \gr /\gl}{\sqrt{g'{}^2 + \gr^2}} \,,
\qquad
\inv{e^2} = \inv{\gr^2} + \inv{ \gl^2} + \inv{g'{}^2} \,.
\eeq
In terms of these fields, the above $3 \times 3$ mass-squared matrix is
reduced to a $2 \times 2$ one, spanning only $(Z,Z')$ with entries
\bea
\mz^2 &=& \frac{e^2}2 \frac{( 1 + \tw^2)^2}{\tw^2} ( v_1^2 + v_2^2 + v_3^2)
\,; \cr
\mzp^2 &=& \frac{e^2}2
\frac{ \gr^4( v_1^2 + v_2^2 + v_3^3 + v_R^2)
 + 2 g'{}^2 \gr^2 ( v_R^2 + 2 v_3^2) + g'{}^4( v_R^2 + 4 v_3^2)}
{( \cw g' g_R)^2} \,;\cr
\dz &=& - \frac{e^2}2 \frac{1 + \tw^2}{\cw\tw g' g_R}
\left[ g_R^2 ( v_1^2 + v_2^2 - v_3^2) - 2 g'{}^2 v_3^2 \right] \,,
\eea
where $\Delta_Z$ is the $Z-Z'$ mixing term.  For the charged vector
bosons, the analogous mass terms are
\bea
\mwl^2 &=& \inv2 \gl^2 ( v_1^2 + v_2^2 + v_3^2) \,; \cr
\mwr^2 &=& \inv2 \gr^2 ( v_1^2 + v_2^2 + v_3^2 + v_R^2) \,; \cr
\dw &=& - \half \gl \gr v_1 v_2 \,.
\eea
Note that the $\rho$ parameter is one at tree-level, i.e. $ \mwl^2 = \cw^2
\mz^2$, in the absence of mixing, i.e. $\Delta_W = \Delta_Z = 0$.
This can be achieved by taking $ v_2 \ll v_1 $ as already discussed in the
text, and requiring
\beq
v_3^2 = \frac{v_1^2 + v_2^2 }{1 + 2 g'{}^2/g_R^2} \simeq \frac{v_1^2 }
{1 + 2 g'{}^2/g_R^2} \equiv \asd^2 v_1^2 ~~
\stackrel{g_L = g_R}\longrightarrow ~~
 (1-2\sin^2 \theta_W)v_1^2  \,.\label{eq:v3}
\eeq
Without this cancellation from $X$, $\dz$ would have been unacceptably
large.

\paragraph{Scalar potential:}
With the addition of $\Chi$, more terms occur in the Higgs potential:
\bea
V_X &=& m_\Chi^2  \tr \cd \Chi +
\qwe_1 \left| \det\Chi \right|^2 +
\qwe_2 \left| \tr \pd \Chi \right|^2 +
\qwe_3 \left| \tr \pt^\dagger \Chi \right|^2 \cr &+&
\qwe_4 \left[ \left( \tr\pd \Chi \right) \left( \tr \cd \pt \right)
+ \hc \right] +
\qwe_5 \left( \tr \cd \Chi \right)^2 +
\qwe_6 \left( \tr \pd \Phi \right) \left( \tr X^\dagger X \right) \cr &+&
\qwe_7 \left[ (\det\Phi) \left( \tr X^\dagger X \right) + \hc \right] +
\qwe_8 \left| \pr \right|^2 \left( \tr X^\dagger X \right) +
\qwe_9 \tr \left( \pd \Chi \cd \Phi \right) \cr &+&
\qwe_{10} \left[ \tr(\pd \Chi \ptd \Chi) + \hc \right] +
\qwe_{11} \left[ \prtd \pd \Chi \pr + \hc \right] +
\qwe_{12} \tr(\pd \Phi \cd \Chi) \cr &+&
\qwe_{13} \tr(\cd \Chi)^2 +
\qwe_{14} \prd \cd \Chi \pr \,,
\eea
where $ \prt = i \sigma_2 \pr^*$. The full potential is then
$ V \to V + V_X $.

The minimum value of $V$, which we denote by $ V_0 $,
occurs when the various neutral fields
are set equal to their corresponding \vev s:
\bea
V_0 &=& m_R^2 v_R^2 + m^2 (v_1^2 + v_2^2) + 2 \mu^2 v_1 v_2 + {1 \over 2}
\lambda_R v_R^4 + {1 \over 2} \lambda_1 (v_1^2+v_2^2)^2 + {1 \over 2} \lambda_2
(v_1^4 + v_2^4) \cr
&+& \lambda_3 v_1^2 v_2^2 + 2 \lambda_4 (v_1^2 +
v_2^2) v_1 v_2 + f_1 v_1^2 v_R^2 + f_2 v_2^2 v_R^2 + 2 f_3 v_1 v_2 v_R^2 \cr
&+& m_\Chi^2 v_3^2 + \qwe_9 v_1^2 v_3^2 + 2 \qwe_7 v_1 v_2 v_3^2 +
\qwe_{12} v_2^2 v_3^2 +
 \qwe_6 (v_1^2 + v_2^2) v_3^2 + \qwe_{13} v_3^4 \cr
&+& f_5 v_3^4 + 2 \qwe_{11} v_1 v_3 v_R^2 +
 \qwe_{14} v_3^2 v_R^2 + \qwe_8 v_3^2 v_R^2 \,,
\eea
where  $v_{R,1,2,3}$ satisfy
\bea
0 &=&
v_1[m^2 + f_1 v_R^2 + (\lambda_1 + \lambda_2) v_1^2 + (\lambda_1 +
\lambda_3) v_2^2 + 3 \lambda_4 v_1 v_2] + v_2 (\mu^2 + f_3 v_R^2 + \lambda_4
v_2^2) \cr
&& + v_3 [ (\qwe_6  + \qwe_9) v_1 v_3 + \qwe_7 v_2 v_3 + \qwe_{11} v_R^2 ] \,; \cr
0 &=&
v_2[m^2 + f_2 v_R^2 + (\lambda_1 + \lambda_2) v_2^2 + (\lambda_1 +
\lambda_3) v_1^2 + 3 \lambda_4 v_1 v_2] + v_1 (\mu^2 + f_3 v_R^2 + \lambda_4
v_1^2) \cr
&&+  v_3 [ (\qwe_6  + \qwe_{12}) v_2 v_3 + \qwe_7 v_1 v_3 ] \,; \cr
0 &=&
v_R (m_R^2 + \lambda_R v_R^2 + f_1 v_1^2 + f_2 v_2^2 + 2 f_3 v_1 v_2)
+ v_3 [ (\qwe_8  + \qwe_{14}) v_R v_3 + 2 \qwe_{11} v_1 v_R ] \,; \cr
0 &=&
v_3[m_\Chi^2 + (\qwe_6 + \qwe_9) v_1^2 + 2 \qwe_7 v_1 v_2 +
(\qwe_{12} + \qwe_6) v_2^2 +
 2 (\qwe_{13} + \qwe_5) v_3^2  \cr
&& + (\qwe_{14} + \qwe_8) v_R^2 ] + \qwe_{11} v_1 v_R^2 \,.
\label{eq:min}
\eea

Let us define
\beq
\begin{array}{ll}
\zxc_1 = f_1 + \asd \qwe_{11} \,, &
\zxc_2 = \qwe_{11} + \asd(\qwe_8 + \qwe_{14}) \,, \cr
\zxc_3 =  \qwe_6 + \qwe_9  + 2 (f_5 + \qwe_{13}) \asd^2  \,,  &
\zxc_4 = \lambda_1 + \lambda_2 + \asd^2 (\qwe_6 + \qwe_9) \,,  \cr
\zxc_5 = \qwe_{12} - \qwe_9 - (2 \qwe_{13} - \qwe_1) \asd^2  \,,   &
\zxc_6 = \asd \qwe_{14} + \qwe_{11} \,.
\end{array}
\eeq
Using this notation, the vacuum expectation values have the following
solution with $v_2 \ll v_1$:
\bea
v_2 &\simeq& \frac{-[\mu^2 + f_3 v_R^2 + (\lambda_4 + \asd^2 \qwe_7)v_1^2] v_1}
{ m^2 + f_2 v_R^2 +
[\lambda_1 + \lambda_3 + \asd^2(\qwe_6 + \qwe_{12})] v_1^2} \,, \cr
&&\cr
v_1^2 &=& \frac{  m_R^2 \zxc_1 - \lambda_R m^2 }
{   \lambda_R \zxc_4 - \zxc_1(\zxc_1 + \asd \zxc_2)} \,, \cr
&& \cr
v_R^2 &=& \frac{-m_R^2 \zxc_4 + m^2 (\zxc_1 + \asd \zxc_2)}
{\lambda_R \zxc_4 - \zxc_1(\zxc_1 + \asd \zxc_2) } \,,
\eea
where $\asd = g_R/\sqrt{g_R^2 + 2g'{}^2}$ was introduced in (\ref{eq:v3}),
and $ v_3 $ is determined by that same equation.
In order for (\ref{eq:min}) to be consistent with (\ref{eq:v3}),
the parameters in the potential must also satisfy
\beq
\asd m_\Chi^2 [ \zxc_1 (\zxc_1 + \asd \zxc_2 ) - \zxc_4  \lambda_R ]
=
m^2 [ \zxc_2 (\zxc_1 +  \asd \zxc_2) - \asd \zxc_3 \lambda_R ]
+
m_R^2 ( \asd \zxc_1 \zxc_3 - \zxc_2 \zxc_4 ) \,.
\eeq
There are many ways to obtain the desired hierarchy,
\begin{equation}
 v_R \gg v_1 \sim v_3 \gg v_2 \, .
\label{eq:hi}
\end{equation}
For example, let $ m_R \gg m,\mu$ and
$ |f_3|, | \zxc_1 | \ll1 $; then
$ v_R^2 \simeq m_R^2/\lambda_R $,
$ v_1^2 \simeq (\zxc_1/\zxc_4) v_R^2 $, and
$v_2 \simeq -(f_3/f_2) v_1 $.

In the limit $v_2=0$, the (unnormalized) would-be Goldstone fields
associated with the $Z\,, Z'\,, W^+_R$ and $W^+_L$ vector bosons are,
respectively,
\bea
 G &=& Im( \ca \phi_1^0 + \sa \chi_2^0) \,; \cr
G' &=& Im \left[ \phi_R^0 -  \sta \epsilon
\left( \asd \phi_1^0 - \chi_2^0 \right)\right]  \,;\cr
G^+_R &=& \phi_R^+ + \epsilon\left( u \chi_1^+ -  \phi_2^+ \right)\,; \cr
G^+_L &=&  \ca \phi_1^+ + \sa \chi_2^+ \,;
\eea
where
\beq
\epsilon = \frac{v_1}{v_R}  \,; \qquad \sa = \sin\alpha,
~~ \ca = \cos\alpha, ~~ \sta = \sin2\alpha ; ~~ \ta = u \,.
\eeq
The physical scalars and their corresponding masses can be obtained from
the potential in a straightforward manner: there is a single doubly-charged
field, 3 singly charged fields and 6 (real) neutral fields. In obtaining
the various expressions we have assumed
(\ref{eq:hi}).
The results are presented in Table~\ref{tab:mass}: they
indicate that the field $ Re( \ca \phi_1^0 + \sa \chi_2^0 ) $
has a mass $ O(v_1 ) $ and plays the role of the SM Higgs boson;
the other physical scalars have masses of order $ v_R $.

\begin{table}
$$
\begin{array}{cc}
{\rm field} & ({\rm mass})^2 \cr
\hline \hline
\chi_1^{++} & - v_R^2 \zxc_6/\asd  \cr \hline
\chi_1^+ & -v_R^2 \zxc_6/\asd \cr
\phi_2^+ & v_R^2 (f_2 -\zxc_1) \cr
(- \sa \phi_1^+ + \ca \chi_2^+) & -2v_R^2  \qwe_{11}/\sta \cr \hline
Re \phi_R^0 & 2 \lambda_R v_R^2 \cr
Im\phi_2^0, ~Re \phi_2^0 & v_R^2 (f_2 -\zxc_1) \cr
Im( -\sa \phi_1^0 + \ca \chi_2^0 ),~
Re( -\sa \phi_1^0 + \ca \chi_2^0 ) & -2 v_R^2 \qwe_{11}/\sta \cr
Re( \ca \phi_1^0 + \sa \chi_2^0 ) & -2 v_1^2 [ (\ca \zxc_1 + \sa \zxc_2)^2/
\lambda_R - ( \sa^2 \zxc_3 + \ca^2 \zxc_4)  ] \cr \hline
\end{array}
 $$
\caption{Physical mass eigenstates and their corresponding
masses in the model containing $\Chi $. We have
ignored corrections of order $ v_1/v_R $ and $ v_2/v_1 $;
the various parameters are constrained by the requirement that all
masses squared must be positive.}
\label{tab:mass}
\end{table}

\newpage
\bibliographystyle{unsrt}

\end{document}